\documentstyle[12pt]{article}

\newcommand{\sect}[1]{\setcounter{equation}{0}\section{#1}}

\newcommand{\EQ}{\begin{equation}}
\newcommand{\EN}{\end{equation}}
\newcommand{\bea}{\begin{eqnarray}}
\newcommand{\ena}{\end{eqnarray}}

\renewcommand{\a}{\alpha}
\renewcommand{\b}{\beta}

\renewcommand{\d}{\delta}

\newcommand{\pa}{\partial}

\newcommand{\p}{\pi}

\renewcommand{\S}{\Sigma}

\begin{document}

\topmargin 0pt
\oddsidemargin 5mm

\renewcommand{\Im}{{\rm Im}\,}
\newcommand{\NP}[1]{Nucl.\ Phys.\ {\bf #1}}
\newcommand{\PL}[1]{Phys.\ Lett.\ {\bf #1}}
\newcommand{\NC}[1]{Nuovo Cimento {\bf #1}}
\newcommand{\CMP}[1]{Comm.\ Math.\ Phys.\ {\bf #1}}
\newcommand{\PR}[1]{Phys.\ Rev.\ {\bf #1}}
\newcommand{\PRL}[1]{Phys.\ Rev.\ Lett.\ {\bf #1}}
\newcommand{\MPL}[1]{Mod.\ Phys.\ Lett.\ {\bf #1}}
\renewcommand{\thefootnote}{\fnsymbol{footnote}}
\newpage
\begin{titlepage}
\vspace{2cm}
\begin{center}
{\bf{{\large CLASSICAL VERSUS QUANTUM SYMMETRIES FOR TODA }}} \\
{\bf{{\large THEORIES WITH A NONTRIVIAL BOUNDARY PERTURBATION}}} \\
\vspace{2cm}
{\large S. Penati} \footnote{E--mail address: penati@mi.infn.it},
{\large A. Refolli} \footnote{E--mail address: refolli@mite35.mi.infn.it}
and {\large D. Zanon} \footnote{E--mail address: zanon@mi.infn.it} \\
\vspace{.2cm}
{\em Dipartimento di Fisica dell' Universit\`{a} di Milano and} \\
{\em INFN, Sezione di Milano, Via Celoria 16, I-20133 Milano, Italy}\\
\end{center}
\vspace{2cm}
\centerline{{\bf{Abstract}}}
\vspace{.5cm}
In this paper we present a detailed study of the quantum conservation
laws for Toda field theories defined on the half plane in the presence
of a boundary perturbation. We show that total
derivative terms added to the currents, while irrelevant at the classical
level,
become important at the quantum level and in general modify significantly the
quantum boundary conservation. We consider the first nontrivial
higher--spin currents for the simply laced $a^{(1)}_n $ Toda theories:
we find that the spin--three current leads to a quantum conserved charge
only if the boundary potential is appropriately redefined through a finite
renormalization. Contrary to the expectation we demonstrate instead
that at spin four the classical
symmetry does not survive quantization and we suspect that this feature
will persist at higher--spin levels. Finally we examine the first nontrivial
conservations at spin four for the $d^{(2)}_3$ and $c^{(1)}_2$
nonsimply laced Toda
theories. In these cases the addition of total derivative terms to
the bulk currents is {\em {necessary}} but sufficient
to ensure the existence of corresponding quantum exact
conserved charges.
\vfill
\noindent
IFUM--522--FT \hfill {December 1995}

\end{titlepage}
\renewcommand{\thefootnote}{\arabic{footnote}}
\setcounter{footnote}{0}
\newpage

\sect{Introduction}

Certain classes of field theories defined on the two--dimensional
plane possess an infinite number of conservation laws which
generalize the energy--momentum conservation. These theories
have been extensively studied primarily because the existence
of such higher--spin conserved charges ensures that their exact
S--matrix can be constructed \cite{b1,b2} and therefore all the on--shell
properties can be determined. In general it has been found \cite{b3,b4} that
whenever the system exhibits a classical symmetry, the corresponding
conservation law can be implemented at the quantum level simply
redefining appropriately the current via the addition of quantum
corrections.
In this way quantum integrability has been established for all affine
Toda theories in their bosonic \cite{b4} as well as in their supersymmetric
\cite{b5} version.
The method which more efficiently allows to construct the
higher--spin currents at the quantum level is massless perturbation
theory. This approach, which is similar in spirit to perturbed conformal
field theory, has the advantage to be applicable
straightforwardly
and to give exact results to all orders in perturbation theory.

Recently there has been much interest in studying these same theories
defined not on the whole plane but on half of it, i.e. on a manifold
with boundary \cite{b6,b7}. In this case one can consider the system in the
presence of a nontrivial boundary perturbation and it turns out
that many physical interesting phenomena can be described in this
fashion \cite{b6,b8}.
A natural question thus arises: how much of the integrability
properties of the original model which locally are still valid in the
interior region, do survive as global symmetries of the theory in the
presence of the boundary? It is easy to show that in order
to construct an integral of motion in terms of the currents conserved in
the bulk, there must be no momentum flowing through the boundary
or at most the momentum evaluated at the boundary has to reduce to
a total time--derivative term. At the classical level this analysis
has been carried out for all Toda theories and it has been shown \cite{b9,b10}
that in general the higher--spin charges are
conserved if the boundary perturbation is chosen appropriately.

These conservation laws if realized at the quantum level
guarantee absence of particle production and the factorization
property of the S matrix. The sine--Gordon model has been studied
quite thoroughly \cite{b7,b11,b12}. One finds
that: i) a renormalization of the currents is sufficient to reabsorb
the quantum anomalies; ii) the most general boundary potential allowed
by the requirement of quantum integrability contains two free
parameters; iii) the exact S--matrix has been constructed and its
connection with the underlying field theory description is
understood.

For the other Toda theories the quantum analysis has not been completed
yet \cite{b13}. Using a generalization of the massless perturbation approach
which is standard for systems without boundary we have started a study
of the quantum boundary conservations for models based on simply laced
\cite{b11} and nonsimply laced \cite{b14} Lie algebras.
In this paper we continue in this task.
In the next section we set our notations and review the
general procedure. In particular we derive the equations that
need be satisfied in order to ensure the existence of a conserved
charge at spin three and spin four level.
In section 3 these results are applied to the specific examples of the
$a^{(1)}_n$ Toda theories. We find that at
spin three a quantum conserved charge exists
if the classical boundary potential is suitably modified
by a finite renormalization.  Instead at spin four, somewhat unexpectedly,
the classical
symmetry does not survive quantization and there are indications
that the charge conservation is broken by true anomalies
at higher--spin levels too. In section 4 we study the first
nontrivial currents (at spin four) for the $d^{(2)}_3$ and $c^{(1)}_2$
nonsimply laced Toda
theories. Here we show that a quantum exact symmetry is
realized only if total derivative terms are added to
the bulk currents. No redefinition of the classical boundary
potential is necessary in these cases. Finally in the last section we draw
our conclusions and make some closing remarks.

\sect{Quantum charge conservation: the general procedure}

We are interested in determining higher conservation laws for Toda--like
systems defined in the upper--half plane. To this end it is convenient
to work in euclidean space with complex coordinates
\EQ
x = \frac{x_0 + ix_1}{\sqrt{2}} \qquad \quad \bar{x} =
\frac{x_0 - ix_1}{\sqrt{2}}
\EN
and corresponding derivatives
\EQ
\pa \equiv \pa_x = \frac{1}{\sqrt{2}} (\pa_0 -i\pa_1) \qquad ~~~
\bar{\pa} \equiv \pa_{\bar x} = \frac{1}{\sqrt{2}} (\pa_0 +i\pa_1) \qquad ~~~
\Box = 2 \pa \bar{\pa}
\label{derivatives}
\EN
We restrict our attention to theories defined by the following action
\EQ
{\cal S} = \frac{1}{\b^2} \int d^2 x \left\{ \theta(x_1) \left[ \frac12
\pa_{\mu} \vec{\phi} \cdot \pa_{\mu} \vec{\phi} + V \right] -
\delta(x_1)B \right\}
\label{action}
\EN
where $B$ denotes the perturbation at the boundary and $V$ is the affine
Toda potential
\EQ
V = \sum_{j=0}^{N} q_j e^{\vec{\a}_j \cdot \vec{\phi}}
\label{1a}
\EN
The Toda theories under consideration are based on a Lie algebra
 ${\cal G}$ of rank $N$ and
we are using the standard notation for the simple roots
$\a_j$, ($j=1,\cdots,N$),
$\a_0 = -\sum_{j=1}^N q_j \a_j$, with $q_j$ the Kac labels ($q_0 = 1$).
{}From the action in (\ref{action}) we can immediately derive
the equations of motion in the bulk region
\EQ
\Box \vec{\phi} =  \sum_{j=0}^N q_j \vec{\a}_j e^{\vec{\a}_j \cdot
\vec{\phi}}
\label{2}
\EN
and at the boundary
\EQ
\left. \frac{\pa \phi_a}{\pa x_1}\right|_{x_1=0} = - \frac{\pa B}{\pa \phi_a}
\label{3}
\EN

It is well known \cite{b15} that for affine Toda theories one can construct an
infinite set of
classical currents and show that they are on--shell conserved in the
interior region $x_1>0$. The classical conservation laws can be written as
\EQ
\bar{\pa} J^{(n)} + \pa \Theta^{(n)} = 0  \qquad \quad
\pa \tilde{J}^{(n)} + \bar{\pa} \tilde{\Theta}^{(n)} = 0
\label{4}
\EN
where $ J^{(n)}$, $\Theta^{(n)}$ denote the two components of a spin $+n$
current, and $\tilde{J}^{(n)}$, $\tilde{\Theta}^{(n)}$ the corresponding
ones for the spin $-n$ current. In terms of these currents one would like
to define a conserved charge and this is feasible if
the following conditions are met at the boundary
\EQ
\left. J_1^{(n)} \right|_{x_1=0} \equiv
\left. i\left( J^{(n)} - \tilde{J}^{(n)} - \Theta^{(n)} + \tilde{\Theta}^{(n)}
\right) \right|_{x_1=0} = -\pa_0 \Sigma_0^{(n)}
\label{9}
\EN
with $\Sigma_0^{(n)}$ a local function of the fields evaluated at $x_1=0$.
Indeed, if this is the case, an integral of motion is given by
\EQ
q^{(n-1)} = \int_0^{+\infty} dx_1 J_0^{(n)} ~~+~ \Sigma_0^{(n)}
\label{8}
\EN
where $J_0^{(n)} = J^{(n)} + \tilde{J}^{(n)} + \Theta^{(n)} +
\tilde{\Theta}^{(n)}$.
It has been shown \cite{b7,b9,b10} that the conditions in eq.(\ref{9})
are satisfied if one restricts the class of boundary potentials to
\EQ
B = \sum_{j=0}^N d_j e^{\frac12 \vec{\a}_j \cdot \vec{\phi}}
\label{1b}
\EN
with appropriate coefficients $d_j$.

We note that equivalent sets of currents can be obtained through
the addition to  $J^{(n)}$, $\Theta^{(n)}$,
$\tilde{J}^{(n)}$, $\tilde{\Theta}^{(n)}$
of total derivative terms
\bea
J^{(n)} \rightarrow J^{(n)}+\pa U~~~~~~~&,&~~~~~~~~
\Theta^{(n)}\rightarrow \Theta^{(n)} -\bar{\pa}U \nonumber \\
\tilde{J}^{(n)} \rightarrow \tilde{J}^{(n)}+\bar{\pa}
\tilde{U}~~~~~~~&,&~~~~~~~~
\tilde{\Theta}^{(n)}\rightarrow \tilde{\Theta}^{(n)} -\pa \tilde{U}
\label{total}
\ena
Indeed the bulk
conservation equations in (\ref{4}) are not modified.
Moreover it is immediate to verify that the charges constructed with the
redefined quantities
\bea
&& J^{(n)}_0 \rightarrow J^{(n)}_0-i \sqrt{2}~\pa_1(U-\tilde{U})~~~~~~~~~~~~~~
J^{(n)}_1 \rightarrow J^{(n)}_1+i \sqrt{2}~\pa_0(U-\tilde{U}) \nonumber\\
&& \Sigma_0^{(n)} \rightarrow \left. \Sigma_0^{(n)}-i \sqrt{2}~
(U-\tilde{U}) \right|_{x_1=0}
\ena
are still conserved and actually coincide with
the ones in (\ref{8}). The net result is that at the classical level
there is the freedom of adding total derivatives
to the currents, but these terms are irrelevant as far as the
conservation laws are concerned. This will not be the case once
quantum corrections will be included.

Thus, in order to proceed further
we discuss now the issue of quantum conservation.
As anticipated in the
introduction the approach best suited to treat the problem exactly,
to all--loop orders, is massless perturbation theory. This method
treats the whole
exponential in (\ref{1a}) as interaction terms without separating
the quadratic parts which would correspond to mass terms. In this way
calculations beyond one loop are much simpler as compared to the
corresponding ones in a massive perturbative approach. We compute
in $x$--space, using for the action in (\ref{action}) massless propagators
defined in the upper--half plane as
\EQ
G_{ij}(x,x') = -\frac{\b^2}{4\pi} \delta_{ij}
\left[ \log{2|x-x'|^2} + \log{2|x-\bar{x}'|}^2 \right]
\label{41}
\EN
and an interaction term
${\cal S}_{i}\equiv {\cal S}_{i}^V+{\cal S}_{i}^B$, where
 ${\cal S}_{i}^V =  \frac{1}{\b^2}
\int_{-\infty}^{+\infty} dx_0 \int_0^{+\infty} dx_1 V$, with $V$ the
affine Toda potential (\ref{1a}) and ${\cal S}^B_{i} = -\frac{1}{\b^2}
\int_{-\infty}^{+\infty} dx_0 B$, with $B$ the boundary perturbation
(\ref{1b}).

At this point
the classical conservation equations in (\ref{4}) and (\ref{9}) can be
reexpressed in perturbation theory as
\EQ
\bar{\pa} \left\langle J^{(n)}(x,\bar{x}) \right\rangle \equiv \bar{\pa}
\left\langle J^{(n)}(x,\bar{x}) ~e^{- {\cal S}_{i}^V} \right\rangle_0 =
{}~-\pa \left\langle \Theta^{(n)} \right\rangle
\label{38}
\EN
for $x_1>0$ and
\EQ
\left. \left\langle J^{(n)}_1(x,\bar{x}) \right\rangle \right|_{x_1=0}
\equiv \left. \left\langle J^{(n)}_1(x,\bar{x}) ~e^{- {\cal S}_{i}}
\right\rangle_0 \right|_{x_1=0} = ~-\pa_0 \left\langle \S_0^{(n)}\right\rangle
\label{39}
\EN
at $x_1=0$. Classical results
correspond to tree level calculations, while quantum corrections are given by
loop contributions.
Normal ordering of the exponentials in $V$ and $B$ is always understood
so that no ultraviolet divergences are produced.

At spin $n$ we consider currents of the form
\EQ
J^{(n)}= \sum c_{ab} \pa^{a_1} \phi_{b_1} \dots \pa^{a_s} \phi_{b_s}
\label{current}
\EN
where $a \equiv (a_1, \cdots , a_s)$ and $\sum a_i=n$.
The coefficients $c_{ab}$ are given in general by a
power expansion in the coupling constant $\b^2$
\EQ
c_{ab}= c^{(0)}_{ab} + \b^2 c^{(1)}_{ab}+\dots
\label{coeff}
\EN
The zeroth order term, $c^{(0)}_{ab}$, must be such that the classical
conservation law in (\ref{4}) is satisfied.

We evaluate quantum corrections in the interior region
Wick contracting the $J^{(n)}$ current with the exponential
in (\ref{38}). Since the current contains terms with at most
$n$ $\pa \phi$ factors, it is clear that we need compute at most up
to $n-1$ loops. Among all the contributions
we want to select the local terms. The ones that are
expressible as total
$\pa$--derivatives contribute directly to the quantum trace in
(\ref{38}); the rest has to vanish if the
current conservation is anomaly free. Therefore one must
determine the yet unknown quantum coefficients in (\ref{coeff})
in order to cancel these potentially anomalous terms. This procedure has been
applied successfully in several examples \cite{b4,b5}.

The actual calculation in (\ref{38})
is simplified
 by the observation that since we are interested only in
local contributions it is sufficient to expand the exponential
to first order in $S_i^V$. Indeed
Wick contractions of the current with the interaction  produce
in general a sum of terms of the form
\EQ
\bar{\pa}_x~ \int d^2 w~ ~{\cal M}(x,\bar{x}) \left[
\frac{1}{(x-w)^k} + \frac{1}{(x-\bar{w})^k} \right] ~{\cal N}(w,\bar{w})
\label{42}
\EN
where ${\cal M}$, ${\cal N}$
 are products of the fields and their
$\pa$--derivatives and the integration is performed in the upper plane.
Local expressions are obtained using in the half plane the relation
\EQ
\bar{\pa}_x \frac{1}{(x-w)^k} = \frac{2\pi}{(k-1)!} \pa_w^{k-1} \delta^{(2)}
(x-w)
\label{43}
\EN
Since only one $\bar{\pa}$ is present, only one interaction factor
(one integration) can appear if we want to obtain a local result.
In this way we
determine the contributions to the quantum trace and the
renormalization of the classical current to all orders
of perturbation theory. This part of the calculation is performed in
the interior
region and in a certain sense it is preliminary and preparatory for the
actual check of the charge conservation in the presence of the
boundary perturbation.

At the boundary $x_1=0$ we have to consider eq.(\ref{39}).
Using the quantum expressions just obtained in the bulk
for $J^{(n)}$ and $\Theta^{(n)}$  we compute
$J_1^{(n)}=i(J^{(n)}-\tilde{J}^{(n)}-\Theta^{(n)}+\tilde{\Theta}^{(n)})$
and then we evaluate its expectation value at $x_1=0$ as in (\ref{39}).
The aim is to isolate local terms which are not $\pa_0$--derivatives
and see if they correspond to true anomalies. In this case the calculation
is complicated by the fact that
local contributions might arise from
higher--order terms in the expansion of the interaction potential,
given now by the complete sum of $V$ in the bulk and $B$ at the
boundary.
Typically expanding the exponential in (\ref{39}) to first
order in ${\cal S}_i^B$ the following structures are produced
\EQ
\lim_{x_1 \to 0}
\int_{-\infty}^{+\infty} dw_0 ~ \left[{\cal P}(x,\bar{x})
 \left( \frac{1}{(x-w)^k}
+ \frac{1}{(x-\bar{w})^k} \right) -\tilde{{\cal P}}(x,\bar{x})
\left( \frac{1}{(\bar{x}-\bar{w})^k} + \frac{1}{(\bar{x}-w)^k} \right)
 \right]{\cal Q}(w_0)
\label{44}
\EN
where ${\cal P}$ and $\tilde{{\cal P}}$ are functions of $\pa^k\phi$
and $\bar{\pa}^k\phi$ respectively. Since $w=\bar{w}$, being $w_1=0$,
the above expression can be written as
\EQ
2(\sqrt{2})^k
{}~\lim_{x_1 \to 0} \int_{-\infty}^{+\infty} dw_0 ~ \left[{\cal P}(x,\bar{x})
\frac{1}{(x_0-w_0 +ix_1)^k}  -\tilde{{\cal P}}(x,\bar{x})
\frac{1}{(x_0-w_0-ix_1)^k} \right]{\cal Q}(w_0)
\label{45}
\EN
Local boundary contributions are obtained selecting in
${\cal P}$ and $\tilde{{\cal P}}$ terms which are equal and
making use of the following relation
\EQ
\lim_{x_1 \to 0^+} \left( \frac{1}{(x_0-w_0 -ix_1)^k} -
\frac{1}{(x_0-w_0+ix_1)^k} \right) = \frac{2 \pi i}{(k-1)!} \pa_{w_0}^{k-1}
\delta^{(1)}(x_0 -w_0)
\label{47}
\EN
Repeating the same procedure, it is clear that Wick contractions
with higher--order factors in the expansion of the boundary
interaction give rise to local contributions whenever the
number of one--dimensional $\d^{(1)}$--functions produced in the
limit $x_1 \rightarrow 0$ equals the number of integrations.

We also have to take into account terms from
the expansion of the bulk potential and/or from mixed factors of
the bulk and the boundary potentials. Such a computation
requires in general a lengthy algebraic
effort. We present an explicit example in Appendix A.

As mentioned above anomalous boundary contributions would correspond
to local terms which cannot be written as $\pa_0$--derivatives of
suitable expressions. Now we want to show that at the quantum
level total derivative terms added to the current and to the
trace might influence these potential anomalies.

The addition of a $\pa U$ term to the $J^{(n)}$ current
modifies the quantum conservation condition in the bulk by
a term $\bar{\pa} \langle \pa U \rangle=
\pa \bar{\pa}\langle U\rangle $.
Obviously, being the result
automatically in the form of a total $\pa$--derivative,
no anomaly is produced in the interior region
and the local terms obtained from $\bar{\pa}\langle U \rangle$
will all contribute to the quantum trace.
Now, while the tree level (classical)
contributions are equal to $\bar{\pa} U$, the loop
(quantum) corrections are not expressible in general as
$\bar{\pa}$--derivatives. Consequently
these terms might lead to  quantum corrections in $J^{(n)}_1$ which
are not $\pa_0$--derivatives and therefore
affect the boundary condition (\ref{39}) in a nontrivial manner.
Since these corrections will play a relevant role in the
following sections, we illustrate this point in detail
with an example.

Let us consider a term to be added to a $J^{(4)}$ current
\EQ
\pa U= c_{ab}\pa(\pa^2\phi_a \pa \phi_b)
\EN
 We start by evaluating $\bar{\pa} \langle \pa U \rangle$
using the massless propagator in (\ref{41}) and dropping
all non--local contributions (with the definition
$\a \equiv \frac{\b^2}{2\pi}$)
\bea
&~&\bar{\pa}\left\langle \pa(\pa^2\phi_a \pa \phi_b) \right\rangle
= \pa \bar{\pa}\left\langle (\pa^2\phi_a \pa \phi_b)
e ^{-\frac{1}{2\pi \a}\int d^2w~V}  \right\rangle_0 \sim \nonumber\\
&~&~~~~~~~~~~~~~~~~~~~~\nonumber \\
&~&~~~~\sim \pa \left\{ -\frac{1}{2\pi\a} \int d^2w \left[
\frac{\a}{2} \pa_x \phi_b ~ \bar{\pa} \left( \frac{1}{(x-w)^2}
+\frac{1}{(x-\bar{w})^2}\right) V_a(w,\bar{w})   +\right. \right.
\nonumber\\
&~& ~~~~~~~~~~~~
\left. \left. -\frac{\a}{2}
\pa^2_x \phi_a ~  \bar{\pa} \left( \frac{1}{x-w}
+\frac{1}{x-\bar{w}}\right) V_b(w,\bar{w})  + \right. \right.
\nonumber\\
&~&~~~~~~~~~~~~\left. \left. -\frac{\a^2}{4}
\bar{\pa} \left( \frac{1}{(x-w)^3}
+\frac{1}{(x-\bar{w})^3} \right) V_{ab}(w,\bar{w}) \right] \right\} \sim
\nonumber\\
&~&~~~~~~~~~~~~~~\nonumber \\
&~&~~~~\sim \pa \left[ \frac{1}{2} \pa V_a \pa \phi_b + \frac{1}{2} V_b \pa^2
\phi_a +\frac{\a}{8} \pa^2 V_{ab}\right]
\ena
Here and in the following we write derivatives of the interactions as
$V_a \equiv \frac{\pa V}{\pa \phi_a}$,
$B_a \equiv \frac{\pa B}{\pa \phi_a}$, and so on.
The contribution to the trace (see eq.(\ref{38})) is then
identified as
\EQ
\Theta \rightarrow -c_{ab}\left[\frac{1}{2} \pa V_a \pa \phi_b +
\frac{1}{2}V_b \pa^2\phi_a
+\frac{\a}{8} \pa^2 V_{ab}\right]
\EN
In the same way one obtains the corresponding contributions
to $\tilde{J}$ and $\tilde{\Theta}$ so that one can compute
the relevant terms produced at the boundary from
$\langle J^{(1)} \rangle$ as in (\ref{39})
\bea
\left\langle J^{(1)} \right\rangle &=& i\left\langle J-\tilde{J}
-\Theta +\tilde{\Theta} \right\rangle \rightarrow
ic_{ab}\left\langle \left[ \pa^3 \phi_a \pa\phi_b + \pa^2\phi_a \pa^2\phi_b
+\frac{1}{2} \pa V_a\pa\phi_b+ \right. \right. \nonumber\\
&~&~~~~~~~~~~~~~\left. \left.+\frac{1}{2} V_b \pa^2\phi_a +
\frac{\a}{8} \pa^2 V_{ab} -c.c. \right] e^{-\frac{1}{2\p\a}
\int d^2w~V +\frac{1}{2\p\a}\int dw_0~B} \right\rangle_0 \nonumber\\
&~&~~~~~~~~~~
\ena
One needs consider terms up to the third--order expansion in $B$
and to second order in the $V$ and $B$ crossed product. We list here
the results from the individual terms and compute in detail the first
one in Appendix A.
\bea
\left. i\left\langle \pa^3\phi_a \pa\phi_b -c.c.\right\rangle \right|_{x_1=0}
&\sim&- 2 B_b \pa^3_0 \phi_a -2 \pa^2_0 B_a \pa_0 \phi_b
+\frac{3}{2} B_b\pa_0 V_a +\frac{1}{2} V_{ac}B_c\pa_0\phi_b \nonumber\\
\left. i\left\langle \pa^2\phi_a \pa^2\phi_b -c.c.\right\rangle \right|_{x_1=0}
&\sim & -2\pa_0 B_a\pa_0^2\phi_b -2\pa_0 B_b\pa^2_0\phi_a
+ V_a \pa_0 B_b +V_b\pa_0 B_a -\frac{2}{3} ~\a \pa_0^3 B_{ab}   \nonumber\\
\left. \frac{i}{2}\left\langle \pa V_a \pa\phi_b -c.c.\right\rangle
\right|_{x_1=0}
&\sim & -\frac{1}{2} B_b \pa_0 V_a
 -\frac{1}{2} V_{ac} B_c\pa_0\phi_b -\a V_{ac} \pa_0 B_{bc} \nonumber\\
\left. \frac{i}{2}\left\langle V_b \pa^2\phi_a -c.c.\right\rangle
\right|_{x_1=0}
&\sim & -V_b \pa_0 B_a \nonumber\\
\left. \frac{i\a}{8}\left\langle  \pa^2V_{ab} -c.c.\right\rangle
\right|_{x_1=0}
&\sim & -\frac{\a}{4} \left[ \pa_0 (V_{abc}B_c)+ \a V_{abcd}
\pa_0 B_{cd} \right]
\label{terms}
\ena
The total sum finally gives
\bea
\langle J^{(1)}\rangle|_{x_1=0} &\sim&c_{ab}\left\{
\pa_0 \left[ -2\pa_0B_a\pa_0\phi_b -2 B_b \pa_0^2\phi_a
-\frac{2}{3} \a \pa_0^2 B_{ab}\right]+ \pa_0\left( V_aB_b\right) +\right.
\nonumber\\
&~&~~~~~ \left. -\a \left[ V_{ac} \pa_0 B_{bc}
+\frac{1}{4}
\pa_0 (V_{abc} B_c) + \frac{\a}{4}
V_{abcd} \pa_0 B_{cd}\right] \right\}
\label{totdev}
\ena
We notice that  contributions containing three $\pa_0$ derivatives,
both classical and quantum,
add up to reconstruct a total $\pa_0$ derivative. Terms containing
one $\pa_0$ derivative behave differently depending on whether they
were produced at tree level or from loops: as expected
the classical terms give rise to
a total $\pa_0$--derivative contribution. The terms instead which
correspond to quantum corrections can modify the boundary condition in
a nontrivial manner and they {\em{must}} be
taken into account while constructing the quantum conserved
charges. A final remark: total derivatives
of the form $ c_a \pa^n \phi_a$ are not relevant since they would
only contribute at the classical level; we will
not consider them in the following.

We turn now to a discussion of the conservation laws for
the specific cases of the spin--3 and spin--4 currents in
Toda theories defined in the upper--half plane, perturbed
by a boundary interaction.

\subsection{Quantum conservation at spin--3 level}

For a Toda theory the action in (\ref{action}) is written in terms
of $n$ independent scalar fields interacting with the potential (\ref{1a})
in the inner region and with a generic perturbation $B$ at the boundary.
According to the general expression in (\ref{current}) we write
a spin--3  current in the form
\EQ
J^{(3)} = \frac{1}{3} a_{abc} \pa \phi_a \pa \phi_b \pa \phi_c +
b_{ab} \pa^2 \phi_a \pa \phi_b +\frac{1}{2} c_{ab} \pa(\pa \phi_a
\pa \phi_b)
\label{69}
\EN
with coefficients $a_{abc}$ and $c_{ab}$
symmetric and $b_{ab}$ antisymmetric in their indices.
We start considering the conservation law of this current
in the upper--half plane following the procedure outlined for the
general case.
We evaluate as in (\ref{38})
$\bar{\pa} \left\langle J^{(3)} \right\rangle$.  We easily find
\bea
&& \bar{\pa} \left\langle J^{(3)} \right\rangle
=\bar{\pa} \left\langle J^{(3)} e^{-\frac{1}{2\pi \a}
\int d^2w~V} \right\rangle_0
\sim \bar{\pa} \left\langle J^{(3)} \left( -\frac{1}{2\pi \a} \right)
\int d^2w ~V \right\rangle_0 \sim\nonumber\\
&&~~~
\sim  \pa \left[ \frac12 b_{ab} V_b \pa \phi_a
+\frac12 c_{ab} V_b \pa \phi_a+\frac{\a}{8} c_{ab} \pa V_{ab}
+ \frac{\a^2}{48}  a_{abc} \pa V_{abc} \right]+
\nonumber \\
&~&~~~ + \frac12 \left[ a_{abc} V_a + 2b_{ac} V_{ab}
+ \frac{\a}{2} a_{abd} V_{acd} \right]
\pa \phi_b \pa \phi_c
\label{70}
\ena
where we have  dropped all the non--local contributions.
Absence of quantum anomalies in the
conservation of $J^{(3)}$ requires that the terms on the right--hand--side,
which are not total $\pa$--derivatives, vanish. This requirement
leads to the following equations for the $a_{abc}$ and $b_{ab}$ coefficients
\EQ
a_{abc}V_a + b_{ac} V_{ab} + b_{ab} V_{ac} + \frac{\a}{4} a_{abd} V_{acd}
+ \frac{\a}{4} a_{acd} V_{abd} =0
\label{71}
\EN
Clearly no restrictions are imposed at this stage on the coefficients
$c_{ab}$ of the total derivative terms.

{}From equation (\ref{70}) we also determine the quantum trace
\EQ
\Theta^{(3)} = -\frac12 b_{ab} V_b \pa \phi_a -
\frac12 c_{ab} V_b \pa \phi_a-\frac{\a}{8} c_{ab} \pa V_{ab}-
\frac{\a^2}{48} a_{abc} \pa V_{abc}
\label{73}
\EN
The same procedure can be applied to compute the quantum currents
$\tilde{J}^{(3)}$, $\tilde{\Theta}^{(3)}$ whose expressions are obtained
from (\ref{69}), (\ref{73}) by exchanging holomorphic derivatives with
antiholomorphic ones.

Now we concentrate on the boundary condition (\ref{39}). Thus we consider
\EQ
\left. \left\langle
i\left(J^{(3)}
-\tilde{J}^{(3)} -\Theta^{(3)} +\tilde{\Theta}^{(3)}\right)
e^{-\frac{1}{2\p\a}
\int d^2w~V +\frac{1}{2\p\a}\int dw_0~B} \right\rangle_0 \right|_{x_1=0}
\EN
Local corrections
come from contractions of the currents with the exponential expanded up
to the third order.  Summing
all the contributions the final result, up to total
$\pa_0$ derivatives, is
\bea
&& \left. \left\langle i(J^{(3)}
-\tilde{J}^{(3)} -\Theta^{(3)} +\tilde{\Theta}^{(3)})
\right\rangle \right|_{x_1=0} \sim \nonumber \\
&&\sim \frac{1}{\sqrt{2}} \left[
\frac{1}{3} a_{abc} B_a B_b B_c + 2b_{ab} V_a B_b
 -\frac{\a}{4} c_{ab}V_{abc}B_c -\frac{\a^2}{24} a_{abc} V_{abcd} B_d
\right]+ \nonumber \\
&& - \frac{1}{\sqrt{2}} \left[ a_{abc} B_a + 4b_{ab} B_{ac}
+2\a a_{abd} B_{acd} \right] \pa_0 \phi_b \pa_0 \phi_c
\label{74}
\ena
In order to cancel the terms proportional to $\pa_0 \phi_b \pa_0 \phi_c$
we require
\EQ
a_{abc} B_a + 2b_{ab} B_{ac} + 2b_{ac} B_{ab} +  \a a_{acd} B_{abd}
+ \a a_{abd} B_{acd} =0
\label{75}
\EN
Comparing (\ref{75}) with (\ref{71}) it is easy to see
that the quantum corrections in both identities
are such that if (\ref{71}) is satisfied with
$V=\sum_{j=0}^n q_j e^{\vec{\a}_j \cdot \vec{\phi}}$, then (\ref{75})
is also satisfied with
$B = \sum_{j=0}^n d_j e^{\frac12 \vec{\a}_j \cdot \vec{\phi}}$.
Finally, once we have found the $a_{abc}$'s and $b_{ab}$'s from
eq.(\ref{71}) or equivalently from (\ref{75}),
we try to determine the coefficients $c_{ab}$ in the current
and the $d_j$'s in the boundary interaction imposing (see again
(\ref{74}))
\EQ
\frac{1}{3} a_{abc} B_a B_b B_c + 2b_{ab} V_a B_b
-\frac{\a}{4} c_{ab}V_{abc}B_c - \frac{\a^2}{24} a_{abc} V_{abcd} B_d =0
\label{76}
\EN
If we are able to satisfy
this condition then we can proceed and construct
the corresponding quantum conserved charge. In section 3 we explicitly
solve the above
equations for the $a^{(1)}_n$ affine Toda theories.

\subsection{Quantum conservation at spin--4 level}

The most general expression for the current of spin four, including
total derivative contributions, has the form
\bea
J^{(4)}&=& \frac{1}{4} a_{abcd} \pa \phi_a \pa \phi_b \pa \phi_c \pa \phi_d +
\frac{1}{2} b_{abc} \pa^2 \phi_a \pa \phi_b \pa \phi_c
+\frac{1}{3} c_{abc} \pa (\pa \phi_a \pa \phi_b \pa \phi_c)+ \nonumber\\
&~&~~~~~~~~~~~~~~~+ \frac{1}{2} d_{ab} \pa^2 \phi_a \pa^2 \phi_b +
e_{ab} \pa(\pa^2 \phi_a \pa \phi_b)
\label{current4}
\ena
with $a_{abcd}$, $c_{abc}$ and $d_{ab}$ completely symmetric,
$b_{abc}$ symmetric in the last two indices and
$b_{abc}+b_{cab}+b_{bca}=0$.
The requirement of
current conservation in the interior region (see eq.(\ref{38}))
determines the quantum trace
\bea
\Theta^{(4)} &=& -\left[ \frac{1}{4} b_{bca} V_a \pa \phi_b \pa \phi_c
+\frac12 c_{abc} V_a \pa \phi_b \pa \phi_c
+\frac{1}{4} d_{ab} \pa V_a \pa \phi_b
+ \frac{1}{2} e_{ab} \pa V_a \pa \phi_b+ \right. \\
&~&~+\frac{1}{2} e_{ab} V_b \pa^2 \phi_a + \frac{\a}{8}
b_{(ab)c} \pa V_{ac} \pa \phi_b +
\frac{\a}{4} c_{abc} \pa V_{ab} \pa \phi_c
+\frac{\a}{48} d_{ab} \pa^2 V_{ab}
+\frac{\a}{8} e_{ab} \pa^2 V_{ab}+
\nonumber\\
&~&~\left. +\frac{\a^2}{32}
a_{abcd} \pa V_{abc} \pa \phi_d +
\frac{\a^2}{48} c_{abc} \pa^2 V_{abc}
+\frac{\a^3}{384} a_{abcd} \pa^2 V_{abcd} \right] \nonumber
\label{trace4}
\ena
and imposes the following two sets of conditions on the coefficients
of the current
\bea
&~& a_{a(bcd)} V_a + \frac{1}{2} b_{a(bc} V_{d)a} -\frac{1}{2} V_{a(d} b_{bc)a}
-\frac{1}{2} d_{a(b} V_{cd)a} + \frac{3}{4} \a a_{ae(cd} V_{b)ae}+
\nonumber\\
&~&~~~~~~~~~+\frac{\a}{8} b_{ae(b} V_{cd)ae} -\frac{\a}{8} V_{ae(cd} b_{b)ae}
+\frac{\a^2}{16} a_{aef(d} V_{bc)aef}=0 \nonumber\\
&~&~~~~~~~~~~~\nonumber \\
&~&b_{[bc]a} V_a + d_{a[b} V_{c]a} + \frac{\a}{4} ( b_{[ba]e}
V_{ace} - b_{[ca]e} V_{abe}) +\frac{\a^2}{8} a_{aef[c} V_{b]aef}
=0
\label{bulk4}
\ena
Then we consider the boundary relation (\ref{39}).
In this case we obtain three sets of equations which
the coefficients and the boundary potential need satisfy in order
to insure the existence of a corresponding quantum charge.
The first two sets arise from terms proportional
to $\pa_0 \phi_b \pa_0 \phi_c \pa_0 \phi_d$ and $\pa^2_0 \phi_b \pa_0 \phi_c$
respectively
\bea
&~& a_{a(bcd)} B_a + b_{a(bc} B_{d)a} -B_{a(d} b_{bc)a}-2d_{a(b} B_{cd)a}
+3\a a_{ae(cd}B_{b)ae}+\a b_{ae(b}B_{cd)ae}+ \nonumber\\
&~&~~~~~~~~~~~~~-\a B_{ae(cd} b_{b)ae} + \a^2 a_{aef(d} B_{bc)aef}=0
\nonumber\\
&~&~~~~~~~~~~~~~\nonumber \\
&~&b_{[bc]a} B_a + 2d_{a[b} B_{c]a} + \a (b_{[ba]e} B_{ace}
-b_{[ca]e} B_{abe}) + \a^2 a_{aef[c} B_{b]aef}=0
\ena
In a way similar to what happened for the spin--3 current, the above equations
do not impose new conditions once the bulk equations (\ref{bulk4})
are satisfied and the boundary potential is of the form
$B = \sum_{j=0}^n d_j e^{\frac12 \vec{\a}_j \cdot \vec{\phi}}$.
The relevant equations are given instead by the terms which are
proportional to $\pa_0 \phi_e$; they must reduce to a total
$\pa_0$--derivative in order to satisfy (\ref{39}). We find
\bea
&~&\left[ a_{abce} B_a B_b B_c + b_{abc} B_b B_c B_{ae} +3\a a_{abcd}
B_b B_c B_{ade} + \right. \nonumber \\
&~&~ + b_{aec} V_a B_c - \frac{1}{2} b_{bea} V_a B_b
- \frac{1}{2} b_{eca} V_a B_c+2d_{ab} V_a B_{be}
-d_{a(b}V_{e)a}B_b+\a b_{abc}V_aB_{bce} +  \nonumber\\
&~&~-\a b_{bca}V_a B_{bce} -\frac{\a}{4} b_{(ab)c} V_{ace} B_b -
\frac{\a}{4} b_{(ae)c} V_{abc}B_b -\a c_{ab(c} V_{e)ab} B_c + \nonumber\\
&~&~-\frac{\a}{12} (d_{ab} +6e_{ab}) V_{abce} B_c-
\frac{\a}{12}(d_{ab}+6e_{ab})V_{abc}B_{ec}-\a
(d_{a(b}V_{c)a}+2e_{a(b}V_{c)a})B_{bce}+
\nonumber\\
&~&~ -\frac{\a^2}{8}a_{acd(e}V_{b)acd} B_b-
\frac{\a^2}{2} b_{(ab)d}V_{acd}B_{bce}-\a^2 c_{abc} V_{abd} B_{cde}+
\nonumber\\
&~&~-\frac{\a^2}{12}(d_{ab}+6e_{ab})V_{abcd}B_{cde}
-\frac{\a^3}{96} a_{abcd}V_{abcdef}B_f -\frac{\a^3}{96} a_{abcd}
V_{abcdf}B_{ef}+  \nonumber\\
&~&~\left.
-\frac{\a^3}{8} a_{adf(c}V_{b)adf} B_{bce} -
\frac{\a^3}{12} c_{abc} V_{abcdf} B_{def}
-\frac{\a^4}{96} a_{abcd} V_{abcdfg}
B_{efg}\right] \pa_0\phi_e  \nonumber\\
&~&~\sim \pa_0-derivative
\label{boundary4}
\ena
In the following sections we attempt to find explicit
solutions of the above equations for specific models.

\sect{The $a^{(1)}_n$ affine Toda theories}

The action for these simply laced theories has the general form
(\ref{action}), (\ref{1a}), with $n$ independent fields
and roots satisfying $\vec{\a}_j^2 = 2$, $\vec{\a}_j \cdot \vec{\a}_k =
-\delta_{j,k\pm 1}$, $j,k=1, \cdots n$, and $q_j=1$, $j=1,\cdots ,n$.

These models possess a  classically conserved spin--3 current of the form
considered in (\ref{69}) \cite{b15}. It has also been established \cite{b4}
that the conservation law in the inner region is valid
at the quantum level. If we call $a^{(0)}_{abc}$ and $b^{(0)}_{ab}$ the
classical coefficients, the quantum solution is given by
\EQ
 a_{abc}=a^{(0)}_{abc} \qquad~~~~,~~~~\qquad b_{ab}=(1+\frac{\a}{2})
b^{(0)}_{ab}
\label{coeff3}
\EN

At spin 3 the boundary condition (\ref{39}) corresponds to equations (\ref{75})
and (\ref{76}). The first set, (\ref{75}), is satisfied
by the coefficients in (\ref{coeff3}) if the boundary potential
is chosen as in (\ref{1b}). Now, in order to satisfy (\ref{76})
which is nonlinear in $B$,
we still have the freedom to choose appropriately the coefficients
$c_{ab}$ in the current and the $d_j$'s in the interaction $B$.
It is convenient to introduce the following definitions
\EQ
a_{ijk} \equiv a_{abc} (\a_i)_a (\a_j)_b (\a_k)_c \qquad
b_{ij} \equiv b_{ab} (\a_i)_a (\a_j)_b   \qquad
c_{ij} \equiv c_{ab} (\a_i)_a (\a_j)_b
\label{78}
\EN
so that from (\ref{71}) we first obtain
\EQ
a_{ijk} + b_{ik} C_{ij} + b_{ij} C_{ik} + \frac{\a}{4} [ a_{iij} C_{ik} +
a_{iik} C_{ij} ] = 0
\label {79}
\EN
where $C_{ij} \equiv \vec{\a}_i \cdot \vec{\a}_j$ is the Cartan
matrix of the $a_n$ Lie algebra.
{}From the above equation and the antisymmetry of $b_{ij}$ one
easily derives
\bea
&& a_{iii} =0 \nonumber\\
&& a_{iij} = - a_{jji} = -\frac{2}{1+\frac{\a}{2}} b_{ij}
\label{80}
\ena
for any $i,j = 0,1,\cdots ,n$. Finally introducing the
notation
\EQ
V = \sum_{j=0}^n e_j^2 \qquad \quad B = \sum_{j=0}^n d_j e_j
\label{77}
\EN
where
\EQ
e_j \equiv e^{\frac12 \vec{\a}_j  \cdot \vec{\phi}}
\label{exponentials}
\EN
we rewrite (\ref{76}) as
\EQ
\sum_{i \neq j} \left( - \frac{1}{4(1+\frac{\a}{2})} d_i^2 +1
\right) b_{ij} d_j  e_i^2e_j -\frac{\a}{8} \sum_{i,j}
c_{ii} C_{ij} d_j e^2_i e_j=0
\label{81}
\EN
It is clear that when in the second sum $i=j$ terms proportional
to $e_i^3$ are produced and the only way to cancel them is to
impose
\EQ
c_{ii}=0
\EN
When $i\neq j$ the coefficients $c_{ij}$ do not enter in (\ref{81}),
so they are undetermined and not relevant.
Consequently, in order to satisfy (\ref{81}) the coefficients
$d_j$ must be chosen as
\EQ
d_j^2 = 4 \left( 1+\frac{\a}{2} \right) \quad , \qquad j=0,1,\cdots ,n
\label{82}
\EN
We note that
setting $\a=0$ the classical result in Ref. \cite{b9,b10} is reproduced, with
$d_j^2 =4$, $j=0,1,\cdots ,n$.
However the presence of quantum corrections
modifies this solution: the conservation of the
$q^{(2)}$ charge requires a nonperturbative, finite
renormalization of the coefficients $d_j$.

\vspace{.5cm}
Now we extend the analysis at spin 4.  The general equations
have been obtained in the previous section, (\ref{current4}),
(2.38),
(\ref{bulk4}), (\ref{boundary4}). We have not been able to
exhibit a solution in the quantum case for a generic $a^{(1)}_n$
Toda system, essentially because the various sets of equations
become highly coupled due to the presence of the perturbative corrections.
So we present the results we have obtained
for the explicit cases $n=3,4,5 $ (a classical spin--4 conserved current
exists only for $n \geq 3$).
Here we discuss  in detail the
$n=3$ example.

The $a^{(1)}_3$ theory is described by three independent
scalar fields. The simple roots can be represented in the
following real form
\EQ
\vec{\a}_1 =(-1,-1,0) \qquad \vec{\a}_2 =(1,0,1) \qquad
\vec{\a}_3 =(-1,1,0)
\EN
so that the bulk potential (\ref{1a}) becomes
\EQ
V=e^{\phi_1-\phi_3}+e^{-\phi_1-\phi_2} + e^{\phi_1+\phi_3}
+ e^{-\phi_1+\phi_2}
\label{Va3}
\EN
and the interaction at the boundary (\ref{1b}) is
\EQ
B= d_0 e^{\frac{1}{2}(\phi_1-\phi_3)}+
d_1 e^{\frac{1}{2}(-\phi_1-\phi_2)} +
d_2 e^{\frac{1}{2}(\phi_1+\phi_3)}
 + d_3 e^{\frac{1}{2}(-\phi_1+\phi_2)}
\label{Ba3}
\EN
The lagrangian is clearly symmetric under
\bea
&~& i)~~~ \phi_2\rightarrow -\phi_2\qquad,\qquad ii)~~~
\phi_3\rightarrow -\phi_3,
\nonumber\\
&~&iii)~~~\phi_1\rightarrow -\phi_1~~~~~~~ \phi_2\rightarrow \phi_3
\label{symmetry}
\ena
In this case it is rather easy to solve the equations
(\ref{bulk4}) and find the coefficients $a_{abc}$,
$b_{abc}$ and $d_{ab}$ which appear
in (\ref{current4}) and (2.38).
The coefficients $c_{abc}$ and $e_{ab}$ are not determined
from the equations in the interior region. Based on the symmetries
in (\ref{symmetry}) the only nonvanishing ones are
\EQ
c_{122}=-c_{133}\equiv G \qquad e_{11}\equiv H \qquad e_{22}= e_{33} \equiv I
\EN
We have
\bea
J^{(4)} &=& (\pa\phi_1)^4+(\pa\phi_2)^4+(\pa\phi_3)^4
-6\left[ (\pa\phi_1)^2(\pa\phi_2)^2+(\pa\phi_1)^2(\pa\phi_3)^2+
(\pa\phi_2)^2(\pa\phi_3)^2\right] + \nonumber\\
&~&~~~-24(1+\frac{\a}{2})\left[ \pa^2\phi_2\pa\phi_2\pa\phi_1
-\pa^2\phi_3\pa\phi_3\pa\phi_1 \right] +(4+\frac{\a^2}{4})
(\pa^2\phi_1)^2 +\nonumber\\
&~&~~~-(8+12\a+\frac{11}{4}\a^2)[(\pa^2\phi_2)^2
+(\pa^2\phi_3)^2]+G\pa\left( \pa\phi_1(\pa\phi_2)^2-\pa\phi_1(\pa\phi_3)^2
\right)+
\nonumber\\
&~&~~~+H\pa(\pa\phi_1\pa^2\phi_1)+I\left[ \pa(\pa\phi_2\pa^2\phi_2)
+\pa(\pa\phi_3\pa^2\phi_3)\right]
\label{current4a3}
\ena
and
\EQ
\Theta^{(4)}= \Theta_1+\Theta_2+\Theta_3+\Theta_0
\label{trace4a3}
\EN
where
\bea
\Theta_1 &=& \left\{ -\left(2+\frac{17}{6}\a+\a^2+\frac{5}{48}\a^3
-\frac{\a^2}{16}G+\frac{\a^2}{8}(H+I)\right)
(\pa\phi_1+\pa\phi_2)^2+ \right.
\nonumber\\
&~&\left.+6\left( 1+\frac{\a}{2}-\frac{1}{12}G\right)(\pa\phi_3)^2+
\left(-\frac{1}{2}H+\frac{\a}{4}G\right) (\pa\phi_1)^2
+\left(\frac12 (1+\a)G-\frac{1}{2}I\right) (\pa\phi_2)^2+\right.\nonumber\\
&~&
+\left((1+\frac{3}{4} \a)G-\frac{1}{2}(H+I)\right) \pa\phi_1\pa\phi_2
+\frac{1}{2}H\pa^2\phi_1+\frac{1}{2}I\pa^2\phi_2+ \nonumber\\
&~&~~~~\left.+
\left( \frac{17}{6}\a+\frac{3}{4}\a^2+\frac{5}{48}\a^3-
\frac{\a^2}{16}G+\frac{\a^2}{8}(H+I)\right)
(\pa^2\phi_1 +\pa^2\phi_2)
\right\} e^{-\phi_1-\phi_2}\nonumber\\
\Theta_2&=&\Theta_1(\phi_1\rightarrow -\phi_1,\phi_2\rightarrow -\phi_3)
\nonumber\\
\Theta_3&=&\Theta_1(\phi_2\rightarrow -\phi_2)\nonumber\\
\Theta_0&=&\Theta_1(\phi_1\rightarrow -\phi_1,\phi_2\rightarrow\phi_3)
\label{trace1234}
\ena
Now we attempt to solve the conditions at the boundary.
As previously emphasized the relevant set of equations
are the ones in (\ref{boundary4}). They are non linear
in $B$, with terms cubic in B and terms which contain the product
$BV$. Using the definitions in (\ref{exponentials})
let us isolate for example terms proportional to $e_0e_1e_2$,
which arise only from
contributions cubic in $B$ (they are not contained in the
products $BV$). Therefore of all the terms in (\ref{boundary4})
we concentrate on the first three and from them we extract
the part proportional to $e_0e_1e_2$. We easily  find that
the classical contributions  cancel and we are left with
\EQ
-\frac{3}{4}\a( \pa_0\phi_1+\pa_0\phi_2)d_0d_1d_2e_0e_1e_2 \equiv
-\frac{3}{4}\a( \pa_0\phi_1+\pa_0\phi_2)d_0d_1d_2e^{\frac{1}{2}(\phi_1-\phi_2)}
\EN
This explicitly shows that there is no way to rewrite this type of
contributions as total $\pa_0$ derivatives
unless we set one of the $d_j, j=0,1,2$, coefficients
equal to zero.
On the other hand since the theory is symmetric under (\ref{symmetry})
the same analysis can be repeated for the terms proportional
to $e_1e_2e_3$, $e_2e_3e_0$, $e_3e_0e_1$. Thus one is forced to set at least
two of the $d_j$'s equal to zero.  In this case it is rather simple to show
that the other nontrivial conditions which follow from (\ref{boundary4})
necessarily require the vanishing of the remaining two $d_j$ coefficients.
As expected the original symmetry of the lagrangian is maintained.
In conclusion the
conservation at the boundary can be implemented only for a
vanishing interaction at the border.

We have repeated the corresponding analysis for the $a^{(1)}_4$
and the $a^{(1)}_5$ Toda theories. These models are described by four
and five scalar fields, respectively. In order to simplify the
algebra we have found convenient to use a realization of
the simple roots of $a_n$ that maintains explicit all the symmetries
of the corresponding affine Dynkin diagram.
This can be achieved
choosing a complex representation of the roots as in Ref. \cite{b2}.
Moreover with that particular choice, as shown in Ref. \cite{b2},
the reality of the lagrangian is implemented representing the
fields in a complex basis with $\phi_a^*=\phi_{n+1-a}$.
It is a simple exercise to modify accordingly the equations
(\ref{boundary4}) which are the relevant ones for checking the
conservation at the boundary. For the specific examples of the
$a^{(1)}_4$
and the $a^{(1)}_5$ Toda systems we have performed
most of the algebraic manipulations using Mathematica.
In both cases we have found that the
classical conservation
of the $q^{(3)}$ charge is broken by quantum anomalous contributions,
following exactly the same pattern as for the $a^{(1)}_3$ theory.
It is from the terms cubic in $B$ that in (\ref{boundary4})
arise contributions
which do not sum up to a total $\pa_0$ derivative. There is no choice of the
coefficients $c_{abc}$ and $e_{ab}$, still undetermined in the $J^{(4)}$ and
$\Theta^{(4)}$ currents, which allows to satisfy the conservation
condition with a nonvanishing boundary potential. We suspect that
a similar situation has to be faced at higher spin levels.

\sect{Conservation laws for nonsimply laced theories}

In this section we study the conservation equations of the first
nontrivial higher--spin current, spin 4, for the two nonsimply laced
$d^{(2)}_3$ and $c^{(1)}_2$ theories. These models, described in
terms of two bosonic fields, are simple enough to allow a complete
analysis. We discuss the two systems separately.

\subsection{The $d^{(2)}_3$ system}

With a realization of the simple roots of the Lie algebra as
\EQ
\vec{\a}_1= (2,0)Ê\qquad \vec{\a}_2= (-1,1)Ê
\EN
one obtains for the  potential in the bulk
\EQ
V= e^{-\phi_1-\phi_2}+e^{2\phi_1}+ e^{-\phi_1 +\phi_2}
\label{Vd2}
\EN
and at the boundary
\EQ
B=d_0~ e^{- \frac12 (\phi_1+\phi_2 )}+d_1~  e^{\phi_1}+
 d_2~ e^{-\frac12 (\phi_1 -\phi_2)}
\label{Bd2}
\EN
This system exhibits the first high--spin conserved current
at spin--4, with a general form given in (\ref{current4}).
Solving the bulk conservation equations (\ref{bulk4})
one finds the coefficients $a_{abcd}$, $b_{abc}$
and $d_{ab}$; the nonvanishing ones are \cite{b4}
\bea
a_{1111} &=& a_{2222}=-\frac{\a}{3}~~~~~~~~~~~~~~~~~
a_{1122} = \frac{2}{3}(1+\frac{\a}{2})~~~~~~~~~
{}~~~~~~~~~b_{221} =2+3\a+\a^2  \nonumber\\
 d_{11}&=& -\frac{\a}{6}(1+3\a+\a^2)~~~~~~~~~~~~~~~~~
d_{22}=2(1+\frac{23}{12}\a+\a^2+\frac{\a^3}{6})
\label{coeffd2}
\ena
The coefficients of the terms which are total derivatives
are not determined; the ones which are allowed by
the symmetry of the lagrangian under $\phi_2\rightarrow -\phi_2$ are
\EQ
c_{122}\equiv G \qquad\qquad c_{111}Ê\equiv 3H \qquad\qquad
e_{11}\equiv I \qquad\qquad e_{22}Ê\equiv J
\EN
Inserting (\ref{coeffd2}) in the general expression (2.38)
for the quantum trace we obtain
\EQ
\Theta^{(4)}=\Theta_0+\Theta_1+\Theta_2
\label{traced2}
\EN
where
\bea
\Theta_0&=&\left\{ \left[ -\frac{\a}{8}\left( 1+\frac{31}{18}\a+\frac{5}{6}\a^2
+\frac{1}{9}\a^3\right) + \frac{\a^2}{16} (G+H) - \frac{\a}{8} (I+J) \right]
(\pa\phi_1+\pa\phi_2)^2+ ~\right. \nonumber\\
{}~&~&~+\left[ \frac{\a}{4} G+
\frac{3}{2}\left( 1+\frac{\a}{2}\right) H -\frac{1}{2}I \right]
(\pa\phi_1)^2~+ \nonumber\\
{}~&~&~+\left[ \frac{1}{2}(1+\a)G -\frac{1}{2} J\right]
(\pa\phi_2)^2+ \nonumber\\
{}~&~&~+\left[ \left( 1+\frac{3}{4}\a\right) G+
\frac{3}{4}\a H -\frac{1}{2} (I+J)\right]
\pa\phi_1 \pa\phi_2+ \nonumber\\
{}~&~&~+\left[ -\frac{\a}{24}\left( 1+\frac{5}{6}\a-\frac{\a^2}{2}
-\frac{\a^3}{3}\right) -\frac{\a^2}{16}(G+H) +\frac{\a}{8} (I+J)\right]
(\pa^2\phi_1+\pa^2\phi_2 )
+\nonumber\\
{}~&~&~+\left. \frac{1}{2} I ~\pa^2\phi_1+ \frac{1}{2} J~ \pa^2\phi_2 \right\}
e^{-\phi_1-\phi_2}\nonumber\\
&~&~~~~~~~~~~\nonumber \\
\Theta_1 &=&\left\{ \left[ -\frac{\a}{6}\left( 1+\frac{10}{3}\a+3\a^2
+\frac{2}{3} \a^3\right)  -(3+6\a+2\a^2)H -2(1+\a)I\right]
(\pa\phi_1)^2+\right.
\nonumber\\
{}~&~&~+\left( 1+\frac{3}{2}\a+\frac12\a^2 -G \right) (\pa\phi_2)^2+
\nonumber\\
{}~&~&~+\left. \left[ -\frac{\a^2}{36}(1+3\a+2\a^2)-\a^2 H
-(1+\a)I\right]
\pa^2\phi_1 \right\} e^{2\phi_1}\nonumber\\
&~&~~~~~~~~~~~~~\nonumber \\
\Theta_2&=& \Theta_0(\phi_2\rightarrow -\phi_2)
\label{trace1d2}
\ena
The requirement of absence of anomalies at the boundary
(\ref{39}) leads to the set of equations in (\ref{boundary4}).
In the specific case under consideration they give
\bea
{}~&~&\frac{\a}{2} (G-3H)d_0=0 \nonumber\\
{}~&~&(3+2\a-\a^2)d_1^2d_0-\left[ 6+10\a+\frac{8}{3}\a^2-\frac{13}{9}\a^3
-\a^4 -\frac{2}{9}\a^5+ \right. \nonumber\\
{}~&~&~~~~~~~~~~~~~~~~~~~~~~~\left. +4\a(3+3\a+\a^2)H+4\a(1+\a)I\right]
d_0=0 \nonumber\\
{}~&~&\frac{\a}{4}d_0^2 d_1+\frac{\a}{2} \left[ 1+\frac{5}{2}\a +\frac{55}{36}
\a^2+\frac{5}{12}\a^3+\frac{\a^4}{18}+(1+\a+\frac{\a^2}{4})G+ \right.
\nonumber\\
{}~&~&~~~~~~~~~~~~~~~~~~~~~~\left.
+(3+3\a+\frac{\a^2}{4})H-(2+\frac{1}{2}\a)I-\frac{1}{2}\a J \right] d_1
=0 \nonumber\\
{}~&~&\frac{\a}{2}(G-3H)d_2=0 \nonumber\\
{}~&~&(3+2\a-\a^2)d_1^2d_2- \left[ 6+10\a+\frac{8}{3}\a^2-\frac{13}{9}\a^3
-\a^4-\frac{2}{9}\a^5+ \right. \nonumber\\
{}~&~&~~~~~~~~~~~~~~~~~~~~~~\left. +4\a(3+3\a+\a^2)H+4\a(1+\a)I\right]
d_2=0 \nonumber\\
{}~&~&\frac{\a}{4}d_2^2 d_1+\frac{\a}{2} \left[ 1+\frac{5}{2}\a +\frac{55}{36}
\a^2+\frac{5}{12}\a^3+\frac{\a^4}{18}+(1+\a+\frac{\a^2}{4})G+ \right.
\nonumber\\
{}~&~&~~~~~~~~~~~~~~~~~~~~~~\left.
+(3+3\a+\frac{\a^2}{4})H-(2+\frac{1}{2}\a)I-\frac{1}{2}\a J \right] d_1
=0
\label{boundaryd2}
\ena
In the limit $\a\rightarrow 0$ all the terms which contain
$G$, $H$, $I$, $J$ vanish and one recovers the classical
boundary equations whose solution fixes the coefficients
$d_j$'s
\bea
&~&a)~~~d_0=d_2=0 \qquad~~~~~~~d_1~~~arbitrary\nonumber\\
&~&b)~~~d_1= \pm \sqrt{2} \qquad~~~~d_0,~d_2~~~arbitrary
\label{classicald2}
\ena
This result is in agreement with Ref. \cite{b10}.

Now we analyze the equations in (\ref{boundaryd2}) at the quantum level.
First we try to find a solution setting to zero all the
terms which are proportional to total derivatives
in the current, i.e. setting $G=H=I=J=0$. The system in (\ref{boundaryd2})
reduces to
\bea
{}~&~&(3+2\a-\a^2)d_1^2d_0-\left[ 6+10\a+\frac{8}{3}\a^2-\frac{13}{9}\a^3
-\a^4 -\frac{2}{9}\a^5\right]
d_0=0 \nonumber\\
{}~&~&\frac{\a}{4}\left[ d_0^2 d_1+ \left( 2+5\a +\frac{55}{18}
\a^2+\frac{5}{6}\a^3+\frac{\a^4}{9} \right) d_1\right]
=0 \nonumber\\
{}~&~&(3+2\a-\a^2)d_1^2d_2- \left[ 6+10\a+\frac{8}{3}\a^2-\frac{13}{9}\a^3
-\a^4-\frac{2}{9}\a^5 \right] d_2=0 \nonumber\\
{}~&~&\frac{\a}{4}\left[d_2^2 d_1+ \left( 2+5\a +\frac{55}{18}
\a^2+\frac{5}{6}\a^3+\frac{\a^4}{9}\right) d_1\right]
=0
\ena
These equations give either the trivial boundary solution
$d_0=d_1=d_2=0$ or
\bea
&~&d_1^2=2+\frac{\frac{2}{9}\a^5 +\a^4
+\frac{13}{9}\a^3-\frac{14}{3}\a^2-6\a}{\a^2-2\a-3} \nonumber\\
&~&d_0^2=d_2^2=-\left[2+5\a+\frac{55}{18}\a^2+\frac{5}{6}\a^3
+\frac{\a^4}{9}\right]
\ena
Clearly these solutions are not acceptable, primarily because
the $d_0$ and $d_2$ coefficients
have imaginary values and therefore the theory does
not appear to be unitary.
Thus we are forced to reconsider the original system in
(\ref{boundaryd2}) with nonvanishing constants $G$, $H$, $I$
and $J$. In this case the solution is not unique. We use this freedom
to set the $d_j$ coefficients equal to their classical values
in (\ref{classicald2}).
In the first case $d_0=d_2=0$, $d_1$ arbitrary but not vanishing,
the equations in (\ref{boundaryd2}) lead to
\bea
&~&1+\frac{5}{2}\a +\frac{55}{36}
\a^2+\frac{5}{12}\a^3+\frac{\a^4}{18}+(1+\a+\frac{\a^2}{4})G+
(3+3\a+\frac{\a^2}{4})H+ \nonumber\\
&~&~~~~~~~~~~~~~~~~~~~~~~~~~
-(2+\frac{1}{2}\a)I-\frac{1}{2}\a J =0
\ena
This condition does not determine the coefficients uniquely.
It is satisfied for example by the non singular
(in the limit $\a\rightarrow 0$) solution
\EQ
H=I=J=0Ê\qquad \qquad G=-\frac{1+\frac{5}{2}\a +\frac{55}{36}
\a^2+\frac{5}{12}\a^3+\frac{\a^4}{18}}{1+\a+\frac{\a^2}{4}}
\EN
We observe that even if the quantum corrections require the presence
of a total derivative term nonvanishing in the classical limit,
this, as already emphasized, does not alter the charge conserved at the
classical level. These perturbative contributions
modify the conservation at the quantum level and are actually
necessary to implement an exact symmetry of the theory.

Exactly the same conclusions can be reached
for the second choice in (\ref{classicald2}), $d_1=\pm \sqrt{2}$,
$d_0$ and $d_2$ arbitrary and non zero. Now the system in (\ref{boundaryd2})
becomes
\bea
&~&3H-G=0 \\
&~&6\a+\frac{14}{3}\a^2-\frac{13}{9}\a^3-\a^4-\frac{2}{9}\a^5
+4\a(1+\a+\frac{\a^2}{3})G+4\a(1+\a)I=0 \nonumber\\
&~&\frac{\a}{4}d_0^2+\frac{\a}{2}\left[1+\frac{5}{2}\a +\frac{55}{36}
\a^2+\frac{5}{12}\a^3+\frac{\a^4}{18}+(2+2\a+\frac{\a^2}{3})G-
(2+\frac{1}{2}\a)I-\frac{1}{2}\a J\right] =0 \nonumber\\
&~&\frac{\a}{4}d_2^2+\frac{\a}{2}\left[1+\frac{5}{2}\a +\frac{55}{36}
\a^2+\frac{5}{12}\a^3+\frac{\a^4}{18}+(2+2\a+\frac{\a^2}{3})G-
(2+\frac{1}{2}\a)I-\frac{1}{2}\a J\right] =0\nonumber
\ena
Therefore, first we need impose a further restriction  $d_0^2=d_2^2$
with respect to the classical result, then we have to solve
the remaining two equations in three unknowns. Again acceptable
solutions exist, not uniquely determined.

\subsection{The $c^{(1)}_2$ system}

We choose the following representation for the simple roots,
$\vec{\a}_1=\sqrt{2}(0,1)$, $\vec{\a}_2=\sqrt{2}(1,-1)$ so that
in terms of two scalar fields the interactions become
\EQ
V= e^{-\sqrt{2}(\phi_1+\phi_2)}+ 2e^{\sqrt{2}\phi_2}+
e^{\sqrt{2}(\phi_1-\phi_2)}
\label{Vc2}
\EN
and
\EQ
B=  d_0 ~e^{- \frac{1}{\sqrt{2}} (\phi_1+\phi_2 )}+
d_1~ e^{\frac{1}{\sqrt{2}}\phi_2}+
d_2 ~ e^{\frac{1}{\sqrt{2}}(\phi_1-\phi_2)}
\label{Bc2}
\EN
The action is symmetric under the exchange $\phi_1\rightarrow -\phi_1$.
Also for this nonsimply laced theory
the first nontrivial high--spin conserved current in the bulk region
is at spin 4. From the general expression in (\ref{current4}) and the
bulk conservation equations (\ref{bulk4}),
we find the coefficients of $J^{(4)}$ (see also Ref. \cite{b4})
\bea
a_{1111} &=& a_{2222}=4~~~~~~~~~~~~~~a_{1122}=-4(1+\a)
{}~~~~~~~~~~~~~~~~b_{112}=-6\sqrt{2}(2+3\a+\a^2)  \nonumber\\
d_{11} &=& -(8+24\a+23\a^2+6\a^3)~~~~~~~~~~~~~~~~
d_{22}=4+6\a+\a^2
\label{coeffc2}
\ena
while $c_{112}\equiv G$, $c_{222}\equiv 3H$, $e_{11}\equiv I$
and $e_{22}\equiv J$ are still undetermined at this stage.
The quantum trace can be computed explicitly using in (2.38)
the values just obtained for the coefficients (\ref{coeffc2})
\EQ
\Theta^{(4)}= \Theta_0+\Theta_1+\Theta_2
\EN
with
\bea
\Theta_0&=&\left\{ \left[-\left(2+\frac{17}{3}\a+\frac{9}{2}\a^2+
\frac{5}{6}\a^3\right)
+\frac{\a^2}{2\sqrt{2}} (G+H) -\frac{\a}{2} (I+J) \right]
(\pa\phi_1+\pa\phi_2)^2+ \right. \nonumber\\
{}~&~&~+\left[\frac{1}{\sqrt{2}}(1 + 2\a)G - I \right]
(\pa\phi_1)^2 +\nonumber\\
{}~&~&~+\left[ \frac{\a}{\sqrt{2}} G+
\frac{3}{\sqrt{2}}(1+\a)H-J \right] (\pa\phi_2)^2 +\nonumber\\
{}~&~&~+\left[\sqrt{2}\left( 1+\frac{3}{2}\a \right) G
+\frac{3}{\sqrt{2}}\a H-I-J \right]
\pa\phi_1 \pa\phi_2 ~+\nonumber\\
{}~&~&~ \left. -\left[\frac{\a}{\sqrt{2}}\left( \frac{1}{3}
-\frac{\a}{2}-\frac{5}{6}\a^2\right) +
\frac{\a^2}{4}(G+H)+\frac{\a}{2\sqrt{2}}(I+J)
\right] (\pa^2\phi_1+\pa^2\phi_2)+ \right. \nonumber \\
{}~&~&+ \left.
\frac{1}{\sqrt{2}}I~\pa^2 \phi_1 + \frac{1}{\sqrt{2}}J~ \pa^2 \phi_2 \right\}
e^{-\sqrt{2}(\phi_1+\phi_2)} \nonumber\\
&~&~~~~~~~~~~~~\nonumber \\
\Theta_1 &=&\left\{ \left[ 6(2+3\a+\a^2)-\sqrt{2}G
\right] \right. (\pa\phi_1)^2+ \nonumber\\
{}~&~&~\left. +\left[ -2\left( 2+\frac{10}{3}\a+\frac{3}{2}\a^2
+\frac{\a^3}{6}\right) -\sqrt{2}\left( 3+3\a+\frac{\a^2}{2}\right) H
-2\left( 1+\frac{\a}{2}\right) J\right] (\pa\phi_2)^2 + \right. \nonumber\\
{}~&~&~-\left.\left[ \frac{\sqrt{2}}{3}\left( \a+\frac{3}{2}\a^2
 +\frac{\a^3}{2}\right) +\frac{\a^2}{2} H+\sqrt{2}
\left( 1+\frac{\a}{2}\right) J\right] \pa^2\phi_2\right\} e^{\sqrt{2}\phi_2}
\nonumber \\
&~&~~~~~~~~~~~\nonumber \\
\Theta_2 &=&\Theta_0 (\phi_1 \rightarrow -\phi_1)
\label{trace2}
\ena
In this case the equations at the boundary (\ref{boundary4})
are
\bea
{}~&~& \a (G-3H)d_0=0 \nonumber\\
{}~&~&(6+\frac{3}{2}\a-\frac{9}{2}\a^2)d_1d_0^2
-[12+23\a+\frac{41}{3}\a^2+\frac{9}{2}\a^3+\frac{5}{6}\a^4+ \\
{}~&~&~~~~~~~~~~~~~~~ -\sqrt{2}\a(3+3\a+\a^2)H+\frac{\a}{2}(1+\a)I+
\frac{\a^2}{2}J] d_1=0 \nonumber\\
{}~&~&3\a^2d_1^2d_0-\a[16+\frac{80}{3}\a+6\a^2-\frac{2}{3}\a^3
-\sqrt{2}(6+6\a+\a^2)H-2(2+\a)I]d_0=0\nonumber\\
{}~&~&\a (G-3H)d_2=0 \nonumber\\
{}~&~&(6+\frac{3}{2}\a-\frac{9}{2}\a^2)d_1d_2^2
-[12+23\a+\frac{41}{3}\a^2+\frac{9}{2}\a^3+\frac{5}{6}\a^4+ \nonumber\\
{}~&~&~~~~~~~~~~ -\sqrt{2}\a(3+3\a+\a^2)H+\frac{\a}{2}(1+\a)I+\frac{\a^2}{2}J]
d_1=0 \nonumber\\
{}~&~&3\a^2d_1^2d_2-\a[16+\frac{80}{3}\a+6\a^2-\frac{2}{3}\a^3
-\sqrt{2}(6+6\a+\a^2)H-2(2+\a)I]d_2=0 \nonumber
\label{boundaryc2}
\ena
It is rather easy to see that the situation is very similar to the
one described in detail for the previous example. First we verify
that in the classical limit $G$, $H$, $I$ and
$J$ do not play any role and we recover the classical
solution for the boundary coefficients
\bea
&~&a)~~~d_1=0 \qquad \quad ~~~~~~~d_0,~d_2~~~arbitrary\nonumber\\
&~&b)~~~d_0=d_2=\pm \sqrt{2} \qquad~~~~d_1~~~arbitrary
\label{classicalc2}
\ena
Then we observe that at the quantum level no consistent solution
can be found without introducing total derivative terms in the
current.
Indeed, setting the total derivatives to zero, the system
in (4.20) becomes
\bea
{}~&~&(6+\frac{3}{2}\a-\frac{9}{2}\a^2)d_1d_0^2
-[12+23\a+\frac{41}{3}\a^2+\frac{9}{2}\a^3+\frac{5}{6}\a^4] d_1=0
\nonumber\\
{}~&~&3\a^2d_1^2d_0-\a[16+\frac{80}{3}\a+6\a^2-\frac{2}{3}\a^3]d_0=0
\nonumber\\
{}~&~&(6+\frac{3}{2}\a-\frac{9}{2}\a^2)d_1d_2^2
-[12+23\a+\frac{41}{3}\a^2+\frac{9}{2}\a^3+\frac{5}{6}\a^4]
d_1=0 \nonumber\\
{}~&~&3\a^2d_1^2d_2-\a[16+\frac{80}{3}\a+6\a^2-\frac{2}{3}\a^3]d_2=0
\ena
The pattern is the same as in the $d^{(2)}_3$ case:
one solution is given by $d_0=d_1=d_2=0$,
that is a vanishing potential at the boundary.  The other
solution corresponds to the following values for the $d_j$ coefficients
\bea
&~&d_1^2=\frac{1}{3\a}(16+\frac{80}{3}\a+6\a^2-
\frac{2}{3}\a^3)\nonumber\\
&~&d_0^2=d_2^2=2+\frac{20\a+\frac{68}{3}\a^2
+\frac{9}{2}\a^3+\frac{5}{6}\a^4}{6+\frac{3}{2}\a-\frac{9}{2}\a^2}
\label{unphysical}
\ena
Again the results in (\ref{unphysical}) are not interesting, in particular
$d_1$ is singular in the classical limit. The way to circumvent the problem is
to include the total derivative terms and reexamine the original set of
equations in (4.20). It is clear that one possibility is to solve
the system with the $d_j$ boundary coefficients  fixed at their
classical values (\ref{classicalc2}). For $d_1=0$, $d_0$ and
$d_2$ arbitrary and non zero, we find
\bea
&~&G-3H=0\nonumber\\
&~&~~~~~~~~~~~~\nonumber\\
&~&16+\frac{80}{3}\a+6\a^2-\frac{2}{3}\a^3-\sqrt{2}(6+6\a+\a^2)H-2(2+\a)I=0
\ena
For $d_0=d_2=\pm \sqrt{2}$, $d_1$ arbitrary and nonzero, (4.20)
gives
\bea
{}~&~& (G-3H)=0 \\
{}~&~& 20\a +\frac{68}{3}\a^2
+\frac{9}{2}\a^3+\frac{5}{6}\a^4+\sqrt{2}\a(3+3\a+\a^2)H-\frac{\a}{2}(1+\a)I-
\frac{\a^2}{2}J =0 \nonumber\\
{}~&~&3\a^2d_1^2-\a[16+\frac{80}{3}\a+6\a^2-\frac{2}{3}\a^3
-\sqrt{2}(6+6\a+\a^2)H-2(2+\a)I]=0 \nonumber
\ena
It is worth emphasizing that in both cases not all the constants
$G,\dots, J$ can be set to zero not even in the classical limit
$\a\rightarrow 0$. The quantum corrections feed back into the
classical results, requiring the presence of nonvanishing total
derivative terms in the spin--4 current.

\sect{Conclusions}

We have studied the quantum properties of higher--spin charges
for affine Toda theories defined on the upper plane in the presence
of a nontrivial perturbation at the border. We have attempted
a systematic analysis of the various theories, but we had to face
the complexity of the algebraic manipulations which an exact quantum
calculation requires.
Moreover the diverse behaviour of different systems and the diverse
behaviour of different spin currents within the same system have
prevented us from completeness. With these caveats, we have accomplished
nonetheless several goals and obtained quite interesting and
unexpected results.

First we have developed a general technique which
allows to address the problem of quantum charge conservation for
a system defined on a manifold with boundary. This method, even if
perturbative in spirit, allows to obtain exact answers, to all loop
orders in perturbation theory. The algebraic difficulty of its
application arises when dealing with higher and higher--spin
currents: however since the procedure is a step by step one it can
be implemented with a computer program.

Second we have tested our approach on several examples, general enough
to illustrate the issues we wished to discuss. Although we have not
found results with a repetitive pattern,
the procedure itself is repetitive and in a sense straightforward to
be applied.

Finally  we have shown that
in the presence of a nonvanishing perturbation at the boundary
the construction of quantum conserved charges is not automatically
guaranteed by the existence of a corresponding classical charge.
At the quantum level total derivative terms added to the currents
become relevant and necessary, in certain cases, for the realization
of  global, exact symmetries. This feature is a peculiar property
of systems defined on manifolds with a non trivial boundary potential.

Perhaps the most striking finding of our study has been the realization
that for the $a^{(1)}_n$ affine Toda theories there is no choice of a
nonvanishing boundary perturbation, no possibility of a quantum
redefinition of the current which allow the quantum existence of certain
higher--spin conserved charges. The first classical conservation spoiled
at the quantum level by anomalies that cannot be cured is at spin 4.
We have checked this failure of the conservation law  on explicit examples
and we expect the same happening at higher spin too.

We have not attempted to repeat this last analysis on nonsimply laced
Toda systems. It would seem that for the $d^{(2)}_3$ and $c^{(1)}_2$
theories similar anomalies might appear at spin $> 4$.

In any case, for all the theories we have considered, at least
one higher--spin
conserved charge has been found and this is sufficient to imply
the existence of factorizable, elastic S--matrices \cite{b16}.
It would be interesting
to proceed in this direction and make precise the correspondence between
S matrices and boundary Toda systems. In particular one would like
to understand  the role played by the {\em quantum} form of the
boundary potential
of the $a^{(1)}_n$ theories in the specific construction of an exact S--matrix.

\appendix\sect{Appendix}

In this Appendix we show with an explicit example how to proceed
in the computations of local contributions to
$\langle J^{(1)}\rangle|_{x_1=0}$.
Let us consider
\EQ
 i\left\langle \left( \pa^3 \phi_a \pa\phi_b
-\bar{\pa}^3 \phi_a \bar{\pa}\phi_b
 \right) e^{-\frac{1}{2\p\a}
\int d^2w~V +\frac{1}{2\p\a}\int dw_0~B} \right\rangle_0
\label{local}
\EN
We want to extract the local terms when this expression is evaluated at
the border $x_1=0$. First, using the definitions in eq. (\ref{derivatives})
we write
\bea
i\left( \pa^3 \phi_a \pa\phi_b
-\bar{\pa}^3 \phi_a \bar{\pa}\phi_b
 \right)&=& \frac{1}{2} \left( -\pa^3_1 \phi_a \pa_0\phi_b
-3\pa_0\pa^2_1\phi_a\pa_1\phi_b \right.\nonumber\\
&~&~~~~~~~~~\left.
 +\pa^3_0\phi_a\pa_1\phi_b
+3\pa^2_0\pa_1\phi_a\pa_0\phi_b\right) \nonumber\\
&=& 2\pa_0^2\pa_1\phi_a\pa_0\phi_b+2 \pa_0^3\phi_a\pa_1\phi_b
\nonumber\\
&~&~~~~~~~~~-3\pa_0\pa\bar{\pa}\phi_a\pa_1\phi_b
-\pa_1\pa\bar{\pa}\phi_a\pa_0\phi_b
\label{mixed}
\ena
In the last equality we have written $\pa_1^2=2\pa \bar{\pa}-\pa_0^2$
so that it is easier to identify in (\ref{local})
the Wick contractions which lead to
local contributions. Indeed we use
\bea
\lim_{x_1 \to 0} \left\langle \pa_1 \phi(x_0,x_1) \phi(w_0,0)\right\rangle_0
&~&= \lim_{x_1 \to 0} ~-\a\left[ \frac{i}{x_0-w_0+ix_1} -\frac{i}{x_0-w_0-ix_1}
\right] \nonumber\\
&~&~~~~~~~~~~~\nonumber \\
&~&= -2\pi \a \d^{(1)}(x_0-w_0)
\label{limit}
\ena
so that we immediately obtain
\EQ
\left. \left\langle 2\pa_0^2\pa_1\phi_a\pa_0\phi_b+2 \pa_0^3\phi_a\pa_1\phi_b
\right\rangle \right|_{x_1=0} \sim -2\pa_0^2B_a\pa_0\phi_b
-2B_b\pa_0^3\phi_a
\EN
For the other two terms in (\ref{mixed})  we need consider
a double expansion in the bulk and boundary potentials.
We have
\bea
&~&\left\langle
\pa_0\pa\bar{\pa}\phi_a\pa_1\phi_b
e^{-\frac{1}{2\p\a}
\int d^2w~V +\frac{1}{2\p\a}\int dw_0'~B} \right\rangle_0
\sim  \nonumber\\
&~&~~~~\sim -\frac{1}{(2\pi \a)^2}
\int d^2w \frac{\pa V}{\pa\phi_a} \left(-\frac{\a}{2}\right) \pa_{x_0}
\pa\bar{\pa}\left( \log{2|x-w|^2} + \log{2|x-\bar{w}|^2}\right)
 \nonumber\\
&~&~~~~~~
\int dw'_0 \frac{\pa B}{\pa\phi_b } (-\a)
\left[ \frac{i}{x_0-w'_0+ix_1}-
\frac{i}{x_0-w'_0-ix_1}\right]
\ena
Using in the upper--half plane
\EQ
\pa\bar{\pa}
\left( \log{2|x-w|^2} + \log{2|x-\bar{w}|^2}\right)=
2\pi \d^{(2)}(x-w)
\EN
and eq.(\ref{limit}) we obtain the result
\EQ
\left\langle
\pa_0\pa\bar{\pa}\phi_a\pa_1\phi_b\right\rangle
\sim -\frac{1}{2} \pa_0 V_a B_b
\EN
The last term is treated in similar manner
\bea
&~&\left\langle \pa_1\pa\bar{\pa}\phi_a\pa_0\phi_b
e^{-\frac{1}{2\p\a}
\int d^2w~V +\frac{1}{2\p\a}\int d
w'_0~B} \right\rangle_0 \sim \nonumber\\
&~&~~~~\sim -\frac{1}{(2\pi \a)^2}\pa_0\phi_b
\int d^2w \frac{\pa^2 V}{\pa\phi_a \pa\phi_c} \int dw'_0
\frac{\pa B}{\pa\phi_c } \left(-\frac{\a}{2}\right)
 \log{2|w-w'_0|^2}  \nonumber\\
&~&~~~~~~~\left(-\frac{\a}{2}\right)
\pa\bar{\pa}\left[ \frac{i}{x_0-w_0+i(x_1-w_1)}+
\frac{i}{x_0-w_0+i(x_1+w_1)}+\right.\nonumber\\
&~&~~~~~~~~~~~~~~~~~~~~~~~~~~~~\left.
-\frac{i}{x_0-w_0-i(x_1-w_1)}
-\frac{i}{x_0-w_0-i(x_1+w_1)}\right]
\ena
Integrating by parts the $\pa\bar{\pa}$ derivatives we obtain a local
contribution when they hit the log and produce a
$\d^{(2)}(w-w_0')$, while the terms in the square bracket, when evaluated
in the limit $x_1 \rightarrow 0$, give rise to a $\d^{(1)}(x_0-w'_0)$.
We then obtain
\EQ
\left\langle \pa_1\pa\bar{\pa}\phi_a\pa_0\phi_b\right\rangle
\sim -\frac{1}{2} V_{ac} B_c \pa_0 \phi_b
\EN
Summing all the contributions the total result quoted in
(\ref{terms}) is recovered.

\vspace{1.0cm}
\noindent
This work has been partially supported by grants no. SC1--CT92--0789 and no.
CEE--CHRX--CT92--0035.

\end{document}